\documentclass[prb,aps,tightenlines,twocolumn,
floatfix]{revtex4}
\usepackage{graphicx}
\usepackage{amssymb}
\usepackage{amsmath}
\usepackage{bm}
\usepackage{colordvi}
\usepackage{mathrsfs}

\begin{document}
\title{Some comments on the note ``Comment on `Single-parameter quantum charge
  and spin pumping in armchair graphene nanoribbons' by Zhu and 
  Berakdar (arXiv:1207.3457)''} 
\author{Y. Zhou}
\author{M. W. Wu}
\thanks{Author to whom correspondence should be addressed}
\email{mwwu@ustc.edu.cn.}
\affiliation{Hefei National Laboratory for Physical Sciences at
  Microscale and Department of Physics, University of Science and
  Technology of China, Hefei, Anhui, 230026, China}

\date{\today}

\begin{abstract}
In response to the comment by Zhu and Berakdar (arXiv:1207.3457)
that the physical system considered by them in their previous
work [Ding {\em et al}., PRB {\bf 84}, 115433 (2011)] is
conceptually different from ours in the appendix of our
latest paper [Zhou and Wu, arXiv:1206.3435], we show why 
they are relevant and further point out conceptually the problems
in their work.  We also point out that the main supporting arguments in their
comment are incorrect.
\end{abstract}

\maketitle
In the latest comment,\cite{Berakdar_comment} Zhu and Berakdar
claimed that the physical system in their
previous paper\cite{Berakdar_11} is {\em conceptually} different from that 
in the appendix of our recent work.\cite{Zhou_pump} In fact, we had 
no intention to comment on the details in
their model or exact calculation, but only wanted to point
 out that the leading effect of
their results is from the artificial cutoff energy which has to be 
taken in their theory.
Evidently, the results which strongly depend on an
artificial cutoff have no physics meaning. We also demonstrated that if treated 
correctly, there is no cutoff energy needed.\cite{Zhou_pump}

In fact, {\it conceptually} the problems in their
work\cite{Berakdar_11} can be simply addressed in the following.
In the absence of the ac field, the transmission, which is proportional to the
differential conductance, always tends to increase with the absolute value of
the energy, thanks to the increase of the density of states. 
This phenomenon can be found in both works\cite{Zhou_pump,Berakdar_09} 
where the effect of the cutoff energy in their model\cite{Berakdar_09}
 is still irrelevant.
After an ac field is applied, the transmission is just the weighted average of
the field-free transmissions corresponding to various sidebands, as shown in 
Eq.~(A.2) in our paper\cite{Zhou_pump} and Eq.~(17) in their
paper.\cite{Berakdar_11} 
Therefore, the transmission still tends to increase with increasing energy.
The artificial cutoff energy is the only possible reason leading to the
decreasing trend in Fig.~2(a) in Ref.~\onlinecite{Berakdar_11}.
In addition, with the increasing ac-field strength, the time-averaged current
with the Fermi energy around the Dirac point increases 
due to the increasing contribution of the sidebands far away from
the Dirac point. However, the current in Fig.~1(b) in
Ref.~\onlinecite{Berakdar_11} decreases with ac-field strength at strong ac
field, which is impossible unless the influence of the artificial
 cutoff energy becomes significant.

We further point out that the main supporting arguments in the
comment by Zhu and Berakdar\cite{Berakdar_comment} are incorrect. 

(1) They claimed that we only discuss the pumping current, i.e., the
time-averaged current with no bias, and hence our investigations have nothing to
do with theirs.\cite{Berakdar_11} In fact, we have addressed
both currents with and without bias. Specifically, in the appendix
addressing the cutoff problem in Ref.~\onlinecite{Berakdar_11}, we focus on
the current with large bias between the ferromagnetic leads, where the current 
is dominated by the bias-driven part.

(2) They claimed that the relation we used
$T_{LR\sigma}^n(\varepsilon)=T_{RL\sigma}^{-n}(\varepsilon+n\Omega)$ is
invalid in the case with ferromagnetic leads due to the broken time-reversal
symmetry, and our approach need to be justified fundamentally in this case. 
However, the statement relating this formula to the time-reversal
symmetry is misleading. The exact condition of this formula is
$H(t)=H(-t)$ in our specific case,
 or more generally $H(t)=H(-t)^\ast$ (note that this is
not equivalent to the time-reversal symmetry in the presence of spin), which
can be proven exactly and has in fact been proven 
in the literature.\cite{Hanggi_rev} Consequently,
this formula is well valid in the case with ferromagnetic leads.

(3) They commented on our definition of the time-averaged current 
\begin{equation}
  \overline{I}={1 \over T_0}\int_0^{T_0}dtI(t)
\end{equation}
and suggested another definition 
\begin{equation}
\overline{I}=\lim_{T\rightarrow\infty}{1 \over T}
\int_{-T/2}^{T/2}dtI(t).
\end{equation}
Nevertheless, these two definitions are exactly equivalent since the current is time
periodic in the time-periodic system in our paper.\cite{Zhou_pump}
Furthermore, they claimed that our definition  leads to the charge accumulation
in the central region. It can be demonstrated to be not true.
In our paper,\cite{Zhou_pump} Eq.~(15) gives the
time-averaged current flowing away from the left lead 
\begin{equation}
  \overline{I}_L=\frac{e}{h}\sum_{\sigma n} \int_{-\infty}^{\infty} 
  {d\varepsilon} [ T_{LR\sigma}^n(\varepsilon)f_L(\varepsilon) 
  -T_{RL\sigma}^n(\varepsilon)f_R(\varepsilon) ].
\end{equation}
Exchanging $L$ and $R$ in the above formula, one obtains
the time-averaged current flowing away from the right lead
\begin{equation}
  \overline{I}_R=\frac{e}{h}\sum_{\sigma n} \int_{-\infty}^{\infty} 
  {d\varepsilon} [ T_{RL\sigma}^n(\varepsilon)f_R(\varepsilon) 
  -T_{LR\sigma}^n(\varepsilon)f_L(\varepsilon) ].
\end{equation}
Thus, $edN_G/dt=\overline{I}_L+\overline{I}_R=0$,
indicating no charge is accumulated in the central region.


\end{document}